\documentclass[11pt,a4paper]{article}
\usepackage[hyperref]{naaclhlt2019}
\usepackage{times}
\usepackage{latexsym}
\usepackage{graphicx}
\usepackage{enumitem}
\usepackage{url}
\usepackage{siunitx} 
\usepackage{calc} 

\aclfinalcopy 

\title{Examining Lower Latency Routing with Overlay Networks}

\author{Aakriti Kedia \\
  {\tt akedia@ucsd.edu} \\\And
  Akhilan Ganesh \\
  {\tt aganesh@ucsd.edu} \\\And
  Aman Aggarwal \\
  {\tt akaggarw@ucsd.edu} \\}

\date{}

\begin{document}
\maketitle

\begin{abstract}
In today's rapidly expanding digital landscape, where access to timely online content is paramount to users, the underlying network infrastructure and latency performance significantly influence the user experience. We present an empirical study of the current Internet's connectivity and the achievable latencies to propose better routing paths if available. Understanding the severity of the non-optimal internet topology with RIPE Atlas stats, we conduct practical experiments to demonstrate that local traffic from the San Diego area to the University of California, San Diego reaches up to Los Angeles before serving responses. We examine the traceroutes and build an experimental overlay network to constrain the San Diego traffic within the city to get better round-trip time latencies.     
\end{abstract}

\section{Introduction}

The current Internet topology is a web of transit and peering connections between numerous autonomous systems, which facilitate the high inter-connectivity of the Internet. However, this inter-connectivity has certain limitations. Since each autonomous system links up with the Internet on the basis of minimum cost, many sub-optimal geographical and physical routing paths are prioritized for the routing of general traffic.

\begin{figure}[h]
  \centering
  \includegraphics[width=0.5\textwidth]{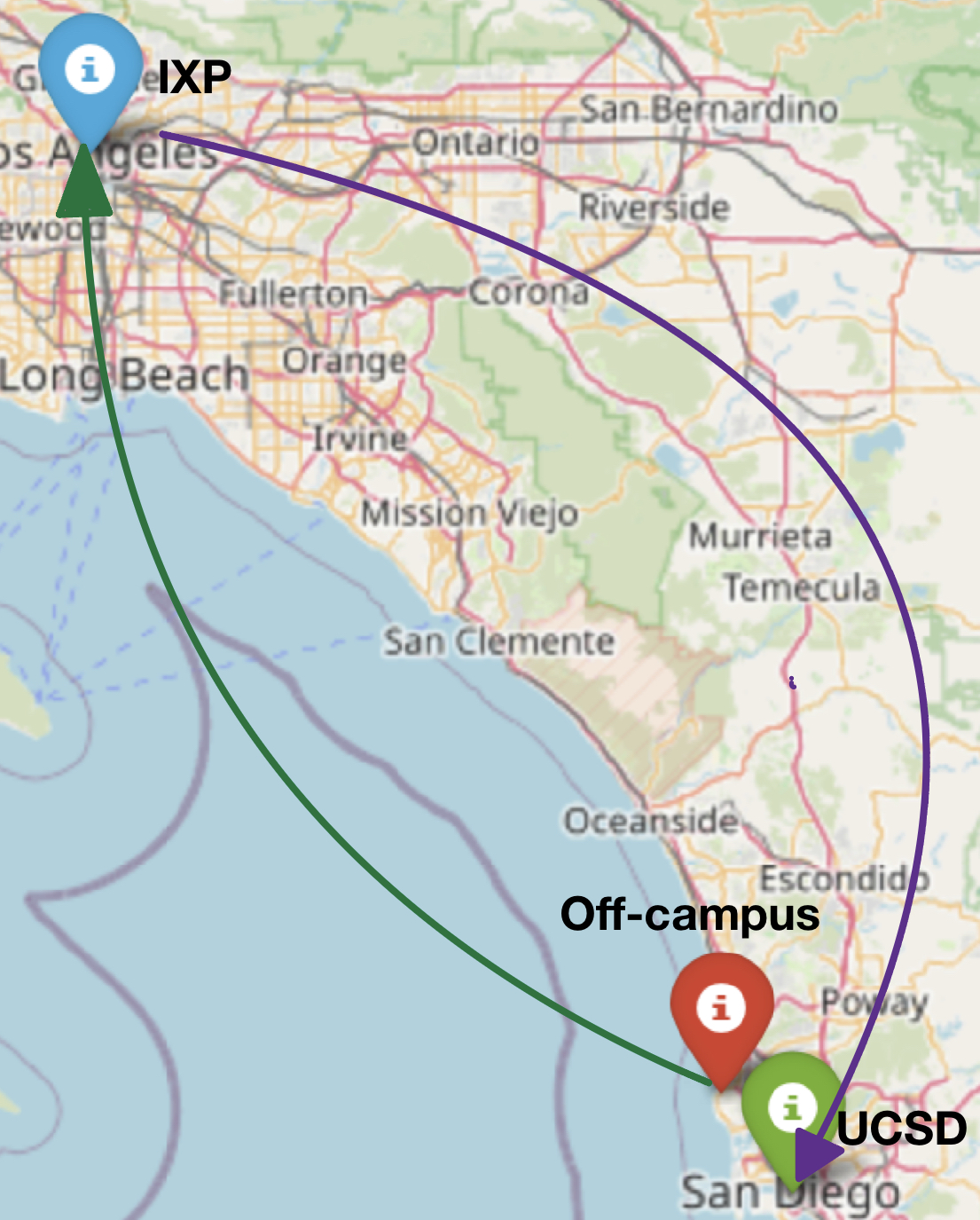}
  \caption{Network path from outside UCSD within San Diego to UCSD ieng6 through LA IXP}
  \label{fig:ac-map}
\end{figure}

The significance of Internet Exchange Points (IXPs) may play a role here. IXPs are physical centers that facilitate peering connectivity between Internet Service Providers (ISPs). Los Angeles (LA) has quite a few IXPs and these are the ones closest to San Diego \cite{datacenter}. This means that any traffic that must be transferred from one ISP to another in San Diego is likely to be routed all the way through LA and back. The connectivity of the University of California, San Diego (UCSD) network with respect to the San Diego area is a direct consequence of this Internet setup. In reality, even while the UCSD network and off-campus ISPs are geographically proximal, packets traveling to and from off-campus users and on-campus servers will still be routed through LA as shown in Figure \ref{fig:ac-map}.

This paper aims to (1) quantify the national severity of non-optimal forwarding paths and disconnectivity, to explore potential lower-latency overlays, (2) explore the topology involved when interacting with UCSD resources from off-campus and confirm the presence of LA IXP in the routing path, and (3) experiment with an overlay network to bypass the LA IXP and reduce effective latency of local UCSD traffic.

\section{Background and Related Work}

An overlay network is a network built on an already existing underlay network infrastructure. The chain of physical or virtual underlay links between two overlay nodes becomes a virtual overlay link on the overlay network \cite{5361643}. Our research explores using overlay networks to gain better performance compared to routing in an underlay network, similar in intent to the role of resilient overlay networks which mitigate network faults \cite{990076}. However, our research explores potential improvements in latency rather than fault tolerance when using overlay networks to exploit or transform the current topology of the Internet.

For our work, we specifically analyze RIPE Atlas data \newcite{ripe-atlas-legal}, and specifically keep in mind the findings of \newcite{10.1145/2815675.2815710} and \newcite{10.1145/2805789.2805796}. As suggested in these papers, we remove noise wherever possible and report only insights with at least $1\%$ improvement. We build on the ideas of efficiency of geographic routing as stated in \newcite{7034368} and our results align with the findings of \newcite{subramanian2002geographic} where packets traveling through different ISPs encounter higher latencies. The criteria and metrics used for reporting results uses the ping and traceroute framework explained in \newcite{paxson1998framework}. We also refer to \newcite{caida-macroscopic} datasets about connectivity and routing of the global Internet.

\section{Methodology}

We quantify the severity of the non-optimal internet topology by rigorously analyzing the publicly available global ping measurement stats provided by RIPE Atlas. We also conduct a UCSD level experiment to examine the layer-3 routing topology and understand the general properties and problems of the underlay network.

\subsection{RIPE Atlas}

We collect the ping stats from the measurement data provided by RIPE for all measurements started after 1st January, 2023 till 19th May, 2023. Data cleanup is performed by filtering only those measurements which are of \textit{stopped} status so that the ongoing and incomplete data do not bias the analysis. 

RIPE has internal probes, which are network devices that are installed to get network stats. Our analysis involved 10000 measurement ids with 11844 global probes involved. We focus primarily on ipv4 ping latencies and keep probes that collect only ipv6 data out of consideration. Under these constraints, we collected 309959 loglines of data and analyze them to create every possible triplet to see if it is possible to redirect data between any source node to a destination node via a middle node but get a better latency compared to the actual ping latency between the source node and the destination node obtained from the measurement data from RIPE. We identify numerous insights about non-optimal routing and internet disconnectivity as detailed in the \ref{Results} section.

To further understand the severity, we further drilled down our analysis to IP level, which is a level deeper than RIPE probes, as a single RIPE probe collects data for groups of ASes. The filtered measurements only include "stopped" status and started after 1st March, 2023. Further, only ipv4 IPs specific to the USA region are focused on to get a more holistic view of the United States region. This leads to a subset of 10,000 measurement ids and around 4212728 loglines of ping data consisting of 16756 unique ips. IP-info online tool \cite{ip-info} was used for mapping IP addresses to the corresponding city, region, and state to further analyze these levels. 

We account for noise in the data by using the median of the ping latency available from the measurements. Each measurement sample from RIPE collects ping latency in three test runs. We sort these runs and move forward with the median of these. Further, data between source-destination pairs collected across different measurements are averaged to minimize the possibility of measurement data errors. We found some interesting insights which are detailed further in \ref{Results}

\subsection{San Diego-UCSD Internet Analysis}

\begin{figure}[h]
  \centering
  \includegraphics[width=0.5\textwidth]{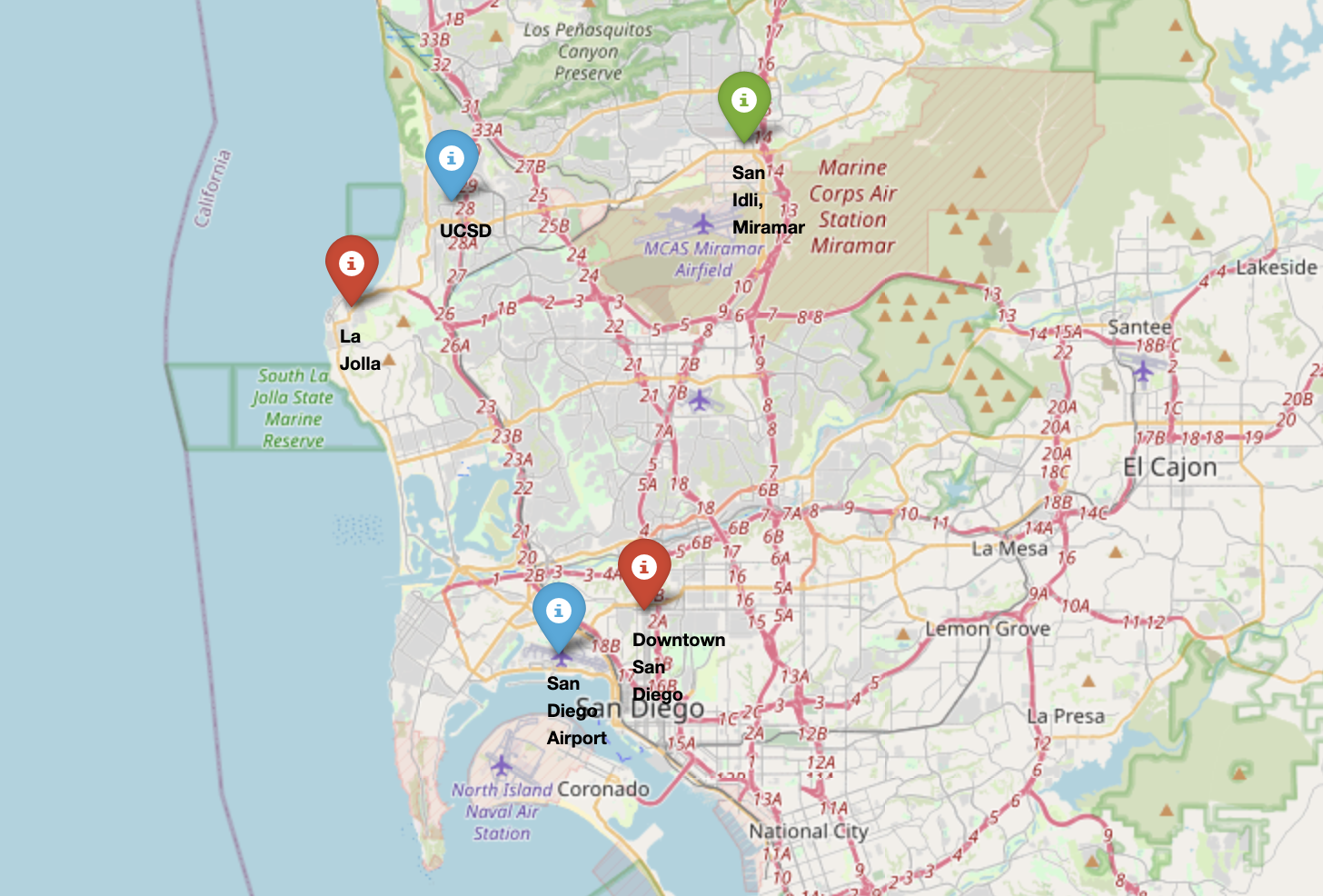}
  \caption{Ping and Traceroute sources to UCSD.}
  \label{fig:san-diego-tour}
\end{figure}

We conduct a preliminary analysis of the Internet topology with respect to the UCSD network. This is done via a hands-on internet measurement data collection experiment of pings and traceroutes to ucsd.edu and ieng6.ucsd.edu from diverse locations in and around the San Diego area and UCSD campus (Figure \ref{fig:san-diego-tour}). These include the following: UCSD CSE building (wifi), UCSD Geisel library (wifi), San Diego downtown (AT\&T wifi, Verizon, and AT\&T), La Jolla downtown (Spectrum wifi and AT\&T), Miramar (Spectrum wifi and AT\&T), and San Diego airport SAN (AT\&T wifi). We exploit both wifi and cellular networks to get more robust results.

We log the ping and traceroute for each pair of source (associated with a specific ISP) and destination. The Time-To-Live field of the ping response indicates the number of hops that a packet takes to arrive at the destination address. We cross-check this value with the number of hops observed from traceroute analyis. Lastly, we record whether sources had packets routed to LA at some point in the traceroute. 

\subsection{Designing and Testing a UCSD Bridge}

\begin{figure}[h]
  \centering
  \includegraphics[width=0.5\textwidth]{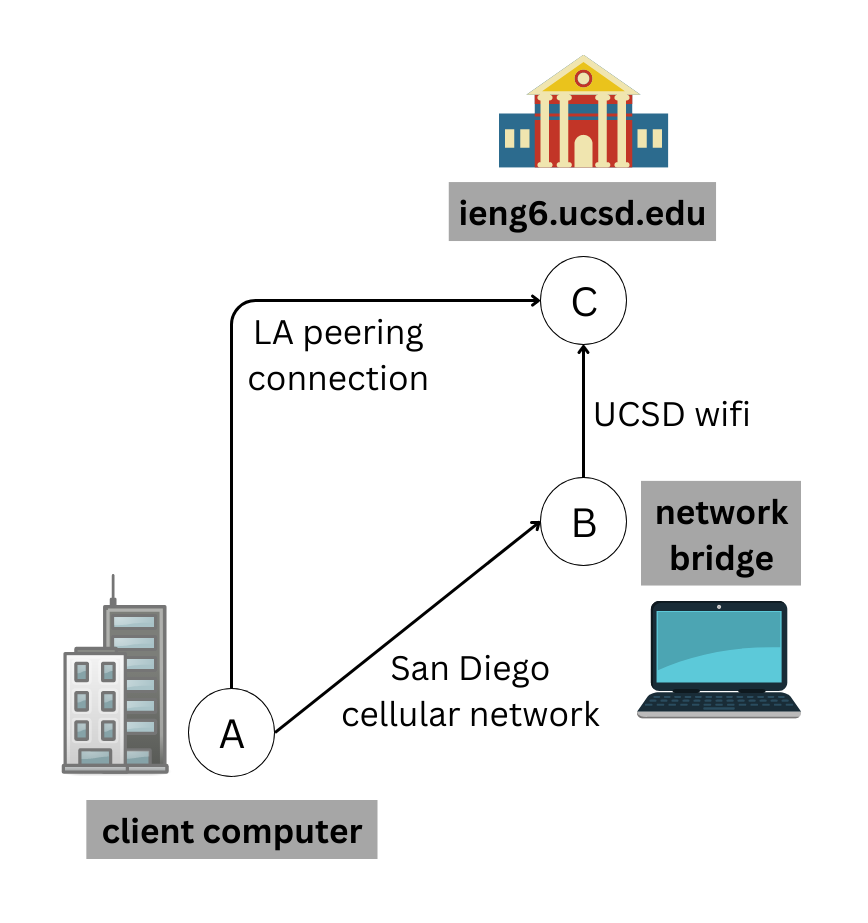}
  \caption{Proposed 3-node overlay network bypassing the LA IXP}
  \label{fig:ABCdiag}
\end{figure}

We propose a 3-hop overlay network solution to route network traffic directly to UCSD without being routed to an LA IXP. The overlay network would resemble the design in Figure \ref{fig:ABCdiag}. The network nodes are as follows:

\textbf{Node A: } A computer connected to a San Diego ISP. This node represents users who want to retrieve UCSD resources (in this case, the ieng6 server) either via the usual layer-3 routing or via the overlay network.

\textbf{Node B: } A forwarding server that is connected on one port to a San Diego ISP and connected on another port to the UCSD network. This node serves as a network bridge that forwards packets through each network, allowing network traffic to potentially bypass the LA IXP.

\textbf{Node C: } The ieng6.ucsd.edu server hosted within the UCSD network. In a full implementation of an overlay network, there could be multiple types of node C, each hosting different resources on the UCSD network.

We attempt to setup a network bridge connected to UCSD wifi and also connected to the AT\&T cellular network. It is posed as an Ethernet connection via USB tethering. We used different ports for the Wifi and cellular network. We plan to receive traffic on the cellular network port which can then be internally forwarded to UCSD Wifi. However, on running the experiment we found that in order to setup a reachable server through cellular we need administrative privileges on the cellular network router. This is to override the Network Address Translation (NAT) system in place and forward traffic to our specific part on the experimental server. Lacking these privileges, we leave an actual implementation of the overlay network and concrete design of node B for future work. Instead, we collect measurements of the RTTs between each node to simulate the expected latencies that such a network is expected to have.

For our simulation, we measure ping times between a device in AT\&T cellular network to another device in the same cellular network. We represent this ping latency as the latency between node A and B. We collect another ping time measurement between a device connected to UCSD wifi to indicate an on-campus resource and an ieng6 server of UCSD, which is another resource within UCSD. This represents the latency between node B and C. For our simulation, we assume that the forwarding time between node B of the cellular network to node B of the on-campus resource is minimal and we don't consider it when reporting results. So in our simulation, A-to-B resides on the AT\&T cellular network, while B-to-C resides on the UCSD network. For each ping measurement, 1000 packets were collected and the round trip times were compared for the analysis. 

Additionally, we analyze the traceroutes of each of these connections to gain a better understanding of the routes that the proposed overlay network takes, and whether it bypasses the LA IXP.

\section{Results}
The following results globally quantify the problem of non-optimality in Internet topology and suggest alternate overlays for better connectivities and latencies.

\subsection{RIPE Atlas data analysis}
\begin{figure}[h]
  \centering
  \includegraphics[width=0.5\textwidth]{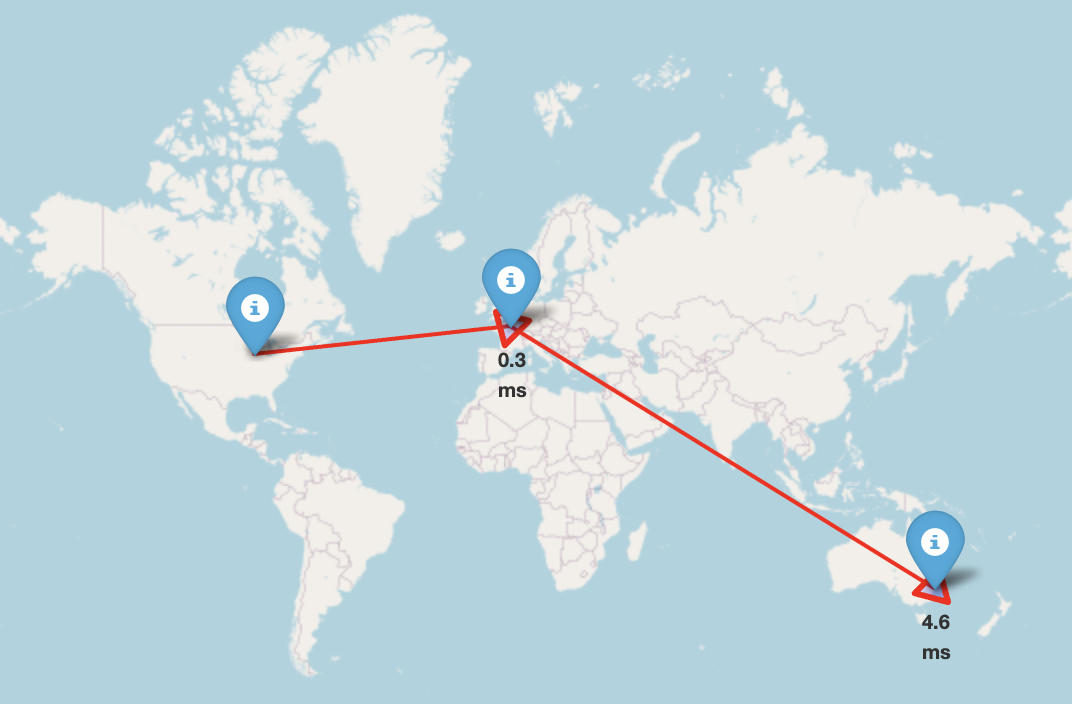}
  \caption{Illinois and Australia via France}
  \label{fig:illinois-france-australia}
\end{figure}

From the experiments conducted on RIPE probes, we identified that Comcast Business Gateway Router Type in the region of Illinois (Probe ID $10194$) 
has no connectivity to New South Wales region of Australia (Probe ID $6636$). We propose an overlay path via France using Bouygues Telecom SA ISP (Probe ID $1003746$). The overlay latency is expected to be less than 5 ms as illustrated in Figure 
 \ref{fig:illinois-france-australia}. Illinois to France (city Paris) takes 0.3 ms and France to Australia 4.6 ms.
 
\label{Results}

\begin{figure}[h]
  \centering
  \includegraphics[width=0.5\textwidth]{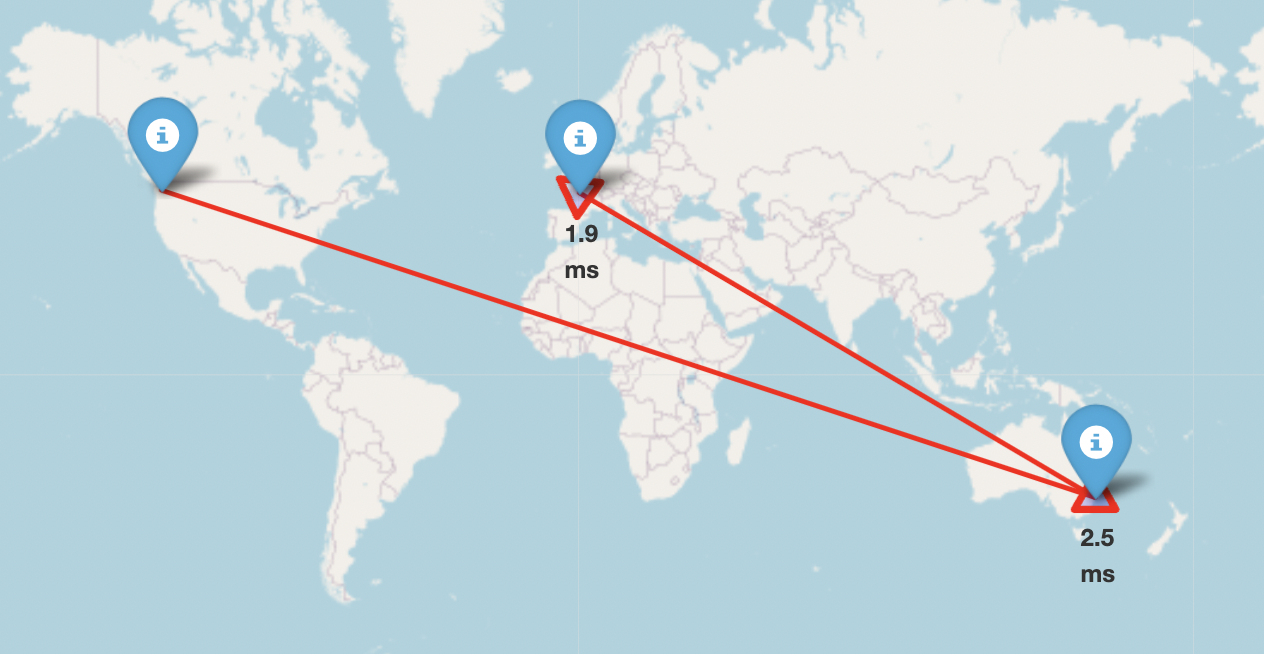}
  \caption{Seattle and France via Australia}
  \label{fig:seattle-france-australia}
\end{figure}

Similarly, Washington region of United States via the EMERALD-ONION ISP (Probe ID $6934$) is connected to Australia (Probe ID $6636$) but lacks connectivity to France (Probe ID $1003746$). An overlay path to France via Australia would be 4.4ms as shown in Figure \ref{fig:seattle-france-australia}.

Another interesting insight is the ping latency from SPACEX-STARLINK ISP, Washington region (Probe ID $62498$) to Australia (Probe ID $6636$). The current round-trip-time is 6.57ms but if forwarded via France using Bouygues Telecom SA ISP (Probe ID $1003746$), an improvement of 0.81ms can be achieved.

\begin{figure}[h]
  \centering
  \includegraphics[width=0.5\textwidth]{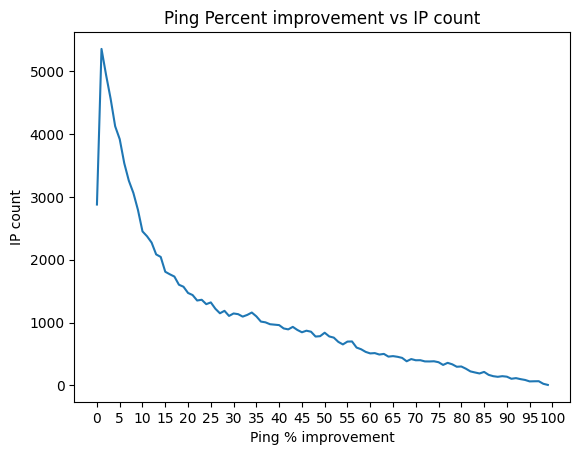}
  \caption{Ping percent improvements vs Source-destination pair counts}
  \label{fig:ping-percent}
\end{figure}

We further drill down on IP level to discover that out of 108356 source IP, destination IP combinations, 105479 combinations could show a ping latency improvement of alteast $1\%$. Table \ref{tab:ping_percent_improvement_table} shows comprehensive quantitative results. Although, a signicant number of combination pairs allow for $<10\%$ of ping latency improvement, there are also combinations which allow upto $100\%$ improvements as depicted by Figure \ref{fig:ping-percent}. 

\begin{table}
\centering
\begin{tabular}{|c|c|}
\hline
\textbf{percent\_improvement} & \textbf{src\_dst\_pairs\_count} \\
\hline
1.0 & 5351 \\\hline
2.0 & 4939 \\\hline
3.0 & 4560 \\\hline
4.0 & 4121 \\\hline
5.0 & 3916 \\\hline
6.0 & 3533 \\\hline
7.0 & 3256 \\\hline
8.0 & 3061 \\\hline
9.0 & 2792 \\\hline
10.0 & 2450 \\
\hline
\end{tabular}

\caption{Ping percent improvements}
\label{tab:ping_percent_improvement_table}
\end{table}

\begin{figure}[h]
  \centering
  \includegraphics[width=0.5\textwidth]{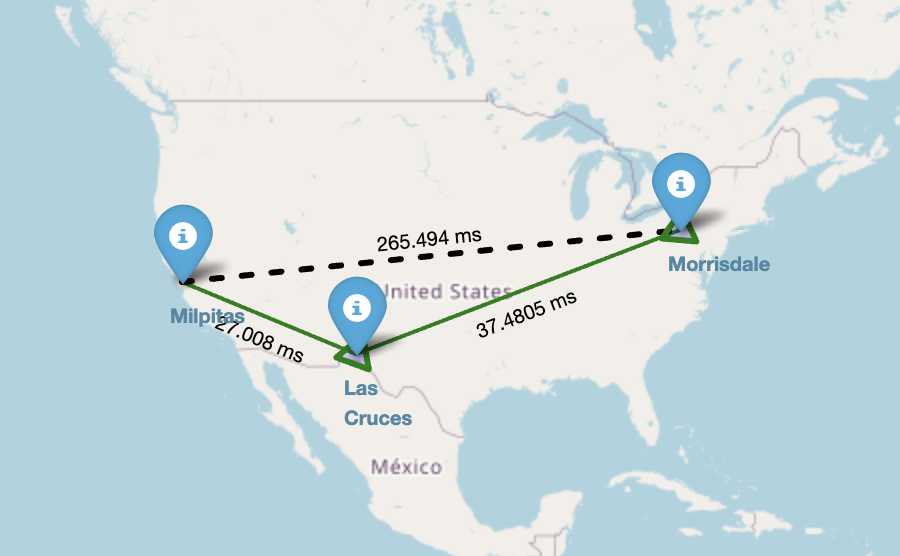}
  \caption{Milpitas and Morrisdale via Las Cruces}
  \label{fig:milpitas-morrisdale}
\end{figure}

One of our insights from IP data includes IPs in Milpitas don't follow an optimal path when directing traffic to IPs in the east US. The current ping time between these IPs go upto as high as 265 ms. Cities like Phoenix, Hilliard, Ashburn, San Francisco can form a lower latency overlay network connecting Milpitas to the east. Figure \ref{fig:milpitas-morrisdale} shows the current ping latency as dashed line (265.49ms), and an optimal path in green of 65ms, giving an improvement of 200ms. As the data is the median of three ping runs and is averaged over all measurement samples, it is reflective of the topology between Milpitas and Morrisdale following a high latency path. 

\begin{figure}[h]
  \centering
  \includegraphics[width=0.5\textwidth]{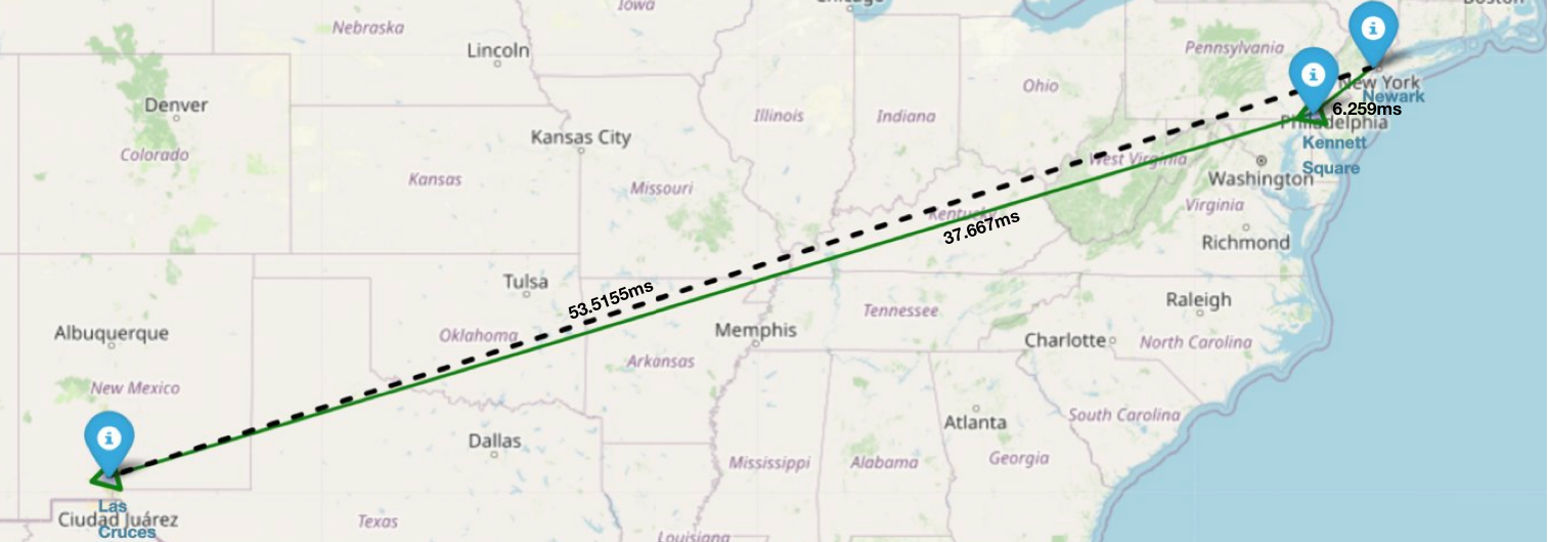}
  \caption{Newark and Las Cruces via Kenett Square}
  \label{fig:milpitas-tulsa}
\end{figure}

Similarly, Figure \ref{fig:milpitas-tulsa} shows another example. Traffic from Newark to Las Cruces takes 53.5ms but when directed through Kenett Square takes 6.25ms from Newark to Kenett Square and 37.6ms from Kenett Square to Las Cruces. 

\begin{figure}[h]
  \centering
  \includegraphics[width=0.5\textwidth]{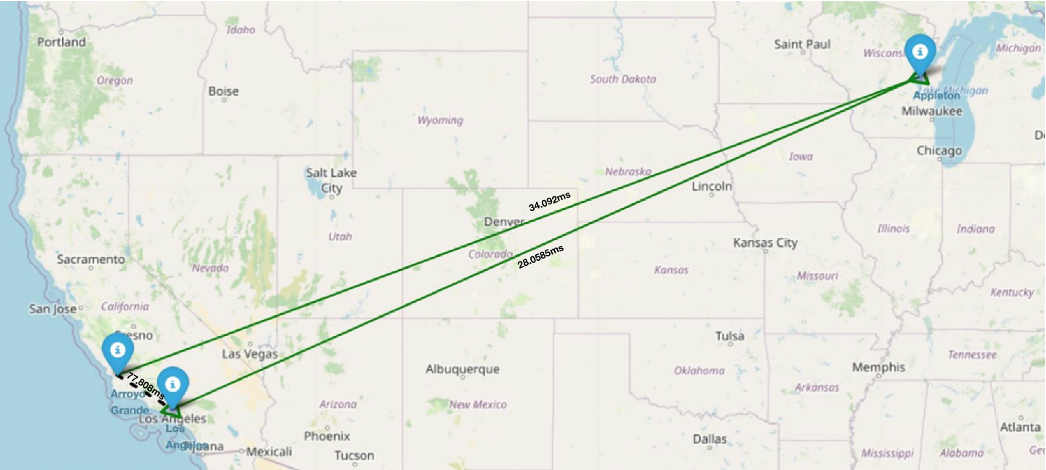}
  \caption{Arroy Grande to LA via Appleton}
  \label{fig:arroy-la}
\end{figure}

However, all internet topology might not necessarily follow shortest paths. Sometimes, distant servers might provide better latencies than those within the same region. For example, Figure \ref{fig:arroy-la} shows that an optimal path of reaching LA
from Arroy Grande hits the east of US all the way to Appleton.
This indicates that some ISPs might have optimal latency servers deployed across geographies or the path latencies might depend on the network congestions.

\subsection{UCSD Network Analysis}

During our traceroute analysis, we figured out that www.ucsd.edu  was externally hosted by Amazon AWS Global Accelerator servers, located in Washington, U.S. Thus, we consider only ieng6.ucsd.edu as the on-campus resource henceforth. Table \ref{tab:ieng6_traceroutes} shows our experimental locations, the number of hops and if the packets are routed through LA IXP. 

Results confirm that almost all external traffic to the ieng6 resource is routed through LA, which the UCSD experimental network bridge aims to address. It further shows that sources within the UCSD network only travel 5 hops to reach the ieng6 server, whereas the outside sources require at least 10 hops.

\begin{table}[h]
  \centering
  \begin{tabular}{|c|c|c|}
    \hline
    \textbf{Experimental source} & \textbf{\# Hops} & \textbf{LA IXP?} \\
    \hline
    UCSD CSE wifi & 5 & No\\
    \hline
    UCSD Geisel wifi & 5 & No\\
    \hline
    SD Downtown wifi & 12 & Yes\\
    \hline
    SD Downtown Verizon & 18 & Yes\\
    \hline
    SD Downtown AT\&T & 12 & Yes\\
    \hline
    LJ Downtown wifi & 14 & Yes\\
    \hline
    LJ Downtown AT\&T & 16 & Yes\\
    \hline
    Miramar wifi & 14 & Yes\\
    \hline
    Miramar AT\&T & 16 & Yes\\
    \hline
    SAN wifi & 15 & Unknown\\
    \hline
  \end{tabular}
  \caption{Number of hops and routing IXP for San Diego locations}
  \label{tab:ieng6_traceroutes}
\end{table}


\newcommand{\abmean}{\num[round-mode=places, round-precision=2]{57.451219512195124}}
\newcommand{\abmedian}{\num[round-mode=places, round-precision=2]{58.0}}
\newcommand{\abvariance}{\num[round-mode=places, round-precision=2]{159.0850188379932}}
\newcommand{\abstd}{\num[round-mode=places, round-precision=2]{12.612890978597777}}
\newcommand{\abmode}{\num[round-mode=places, round-precision=2]{59}}

\newcommand{\bcmean}{\num[round-mode=places, round-precision=2]{10.466880566801635}}
\newcommand{\bcmedian}{\num[round-mode=places, round-precision=2]{12.216999999999999}}
\newcommand{\bcvariance}{\num[round-mode=places, round-precision=2]{11.510906220553505}}
\newcommand{\bcstd}{\num[round-mode=places, round-precision=2]{3.392772644984262}}
\newcommand{\bcmode}{\num[round-mode=places, round-precision=2]{12.52}}

\newcommand{\sumacmean}{\num[round-mode=places, round-precision=2]{67.918100079}}
\newcommand{\sumacmedian}{\num[round-mode=places, round-precision=2]{70.217}}
\newcommand{\sumacvariance}{\num[round-mode=places, round-precision=2]{170.5959250585}}
\newcommand{\sumacstd}{\num[round-mode=places, round-precision=2]{13.061237501}}
\newcommand{\sumacmode}{\num[round-mode=places, round-precision=2]{71.52}}

\newcommand{\acmean}{\num[round-mode=places, round-precision=2]{61.723446893787575}}
\newcommand{\acmedian}{\num[round-mode=places, round-precision=2]{62.0}}
\newcommand{\acvariance}{\num[round-mode=places, round-precision=2]{117.50267669607766}}
\newcommand{\acstd}{\num[round-mode=places, round-precision=2]{10.839865160419555}}
\newcommand{\acmode}{\num[round-mode=places, round-precision=2]{61}}

\begin{figure}[h]
  \centering
  \includegraphics[width=0.5\textwidth]{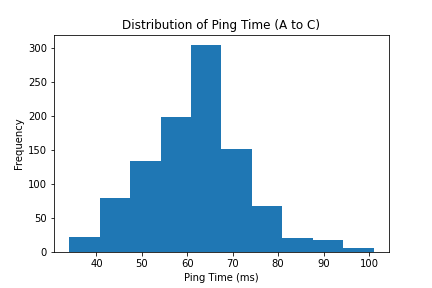}
  \caption{Unimodal RTT frequency distribution from A to C}
  \label{fig:ping-AC}
\end{figure}

\begin{figure}[h]
  \centering
  \includegraphics[width=0.5\textwidth]{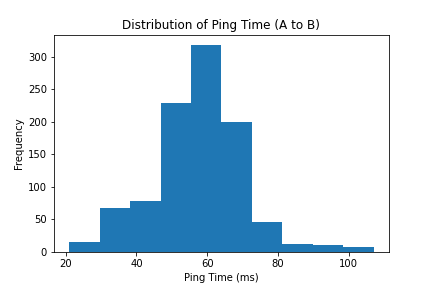}
  \caption{Unimodal RTT frequency distribution from A to B}
  \label{fig:ping-AB}
\end{figure}

\begin{figure}[h]
  \centering
  \includegraphics[width=0.5\textwidth]{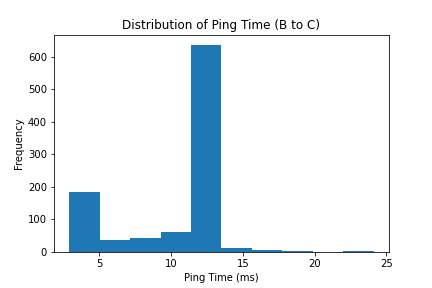}
  \caption{Bimodal RTT frequency distribution from B to C}
  \label{fig:ping-BC}
\end{figure}

Experimental results from the simulated overlay network are demonstrated in Figure \ref{fig:ping-AB} (latency between node A and B), Figure \ref{fig:ping-BC} (latency between node B and C), and Figure \ref{fig:ping-AC} (latency between node C and A). We consider ping latencies as continuous values and plot the distributions for the measurements between all pairs of nodes.

\begin{table*}[htbp]
    \centering
    \begin{tabular}{|l|c|c|c|c|c|}
        \hline
        & Mean (ms) & Median (ms) & Variance (ms$^2$) & Mode (ms) & Std Dev. (ms) \\
        \hline
        Ping between A and B & \abmean & \abmedian & \abvariance & \abmode & \abstd \\
        \hline
        Ping between B and C & \bcmean & \bcmedian & \bcvariance & \bcmode & \bcstd \\
        \hline
         Ping between A and C via B & \sumacmean & \sumacmedian & \sumacvariance & \sumacmode & \sumacstd \\
        \hline
        Ping between A and C & \acmean & \acmedian & \acvariance & \acmode & \acstd \\
        \hline
    \end{tabular}
    \caption{Summary statistics of ping time data for all the three routes - A and B, B and C, and A and C - and aggregate of data from A and B and B and C to predict that between A and C via B}
\end{table*}

As illustrated in Figure \ref{fig:ping-AC}, the ping latencies between A and C follow a unimodal distribution with an average ping latency of $\acmean$ and a median of $\acmedian$. These high statistics are indicative of the traceroute results of traffic being routed from A to LA IXP and then down to C as depicted in Figure \ref{fig:ac-map}. It's interesting to note that the standard deviation in this analysis is moderate (\~16\%) which can indicate some congestion in the network. 

Figure \ref{fig:ping-AB} shows that the ping times distribution from A to B is also unimodal but with a lower mean $\abmean$ and median $\abmedian$ as compared to route A to C. Traceroute analysis from A to B shows that the packets are served within the San Diego network and hence have a round trip time latency of $4ms$ lower compared to A to C. The difference $4ms$ is still underwhelming as the route covers a lesser geographical distance (within San Diego vs through LA). The standard deviation for this sample is $\abstd$ which indicates an instability in the network which can be because of routing delays like queuing delay, propagation delay, etc.  

As Figure \ref{fig:ping-BC} explains, distribution for ping times between B and C takes on a bimodal distribution with first mode at $4ms$ and the second one at \bcmode, with the second mode being significantly more frequent than the first. As it is a bimodel distribution, the mean $\bcmean$ is not a good representation of the central location of the data as it can be affected by the two modes \label{bimodal-mean-invalid}. Therefore, we rely on median ($\bcmedian$ ms), a better estimate of the central tendency of the data. The standard deviation for this distribution is $\bcstd$ ms which is comparatively lower than the other two routes. This result aligns with our expectations that route B to C should show better latencies as it is within the internal UCSD network and can be transmitted with lesser number of hops. Standard deviation of \bcstd ms with respect to the median of \bcmedian ms still conveys network instability.

We combine the statistics of ping latencies from A to B with those from B to C to represent the estimated latencies from A to C via an overlay through B. For standard deviation, we add the two variances and then square-root the sum to resemble the combined deviation. For all other metrics, we add the results from A to B and from B to C.

Due to the bimodal distribution of ping times from B to C, mean is not a good statistic for comparison \ref{bimodal-mean-invalid}. Therefore, we consider the mode or median to compare the ping times from A to C directly versus through node B. Comparing the medians, we observe that ping latency from A to C directly would be $8.22$ ms faster compared to the latency through B. Similarly, comparing the modes reveals that latency from A to C directly is $10.52$ms faster than through B. 

Though the results did not align with RIPE Atlas results of strategically designed overlay networks yielding better latencies, future work in this direction can still yield promising results. As our sample data was restricted to only AT\&T and UCSD ASes, the benefits of the overlay network were not prominent. We further discovered that an AT\&T router on the path from A to B was located in LA, and therefore traffic was unavoidably being routed through LA, which forfeits our assumption of the traffic between A and B being constrained within San Diego. Additionally, as ASes are financially incentivized to optimize for shortest-path routing, the simulated setup might be able to propose improvements only when multiple ASes are involved and transit relationships influence path forwarding. 

\section{Conclusion}

Having rigorously analyzed the network topology using RIPE Atlas measurements, we found great insights about the connectivity and the non-optimal latencies between geographies. We further simulated a hands-on experiment to demonstrate that local San Diego traffic is being routed to LA. Despite simulating the experiment, we could not achieve an improvement in routing the local traffic within San Diego due to lack of network administrative privileges and mobile ISP routing going through LA. The RIPE Atlas results are promising and we still believe that constraining traffic within San Diego would improve latencies because UCSD on-campus pings to ieng6 yield much lower latencies than off-campus pings. As future work, we aim to conduct the same experiment on a larger scale with at least a million data points and a server that is directly reachable within San Diego ISPs and that can also access the UCSD network directly. We further consider wireless networks as a possibility of network instability and hence conducting the experiment on wired systems can also prove fruitful.

\footnotesize
\bibliography{yourbib}


\end{document}